\documentstyle[12pt,psfig]{article}

\newcommand{\eq}[1]{(\ref{#1})}
\newcommand{\ab}{( \alpha - \beta )} 
\newcommand{\nn}{\nonumber}
\newcommand{\fr}{\frac}
\newcommand{\mh}{m_{H^0}^2}
\newcommand{\mH}{m_{h^0}^2}
\newcommand{\mG}{m_{H^{\pm}}^2}
\newcommand{\mA}{m_{A^0}^2}

\newcommand{\mw}{m_W^2}
\newcommand{\mz}{m_Z^2}

\newcommand{\al}{\alpha}
\newcommand{\be}{\beta}
\newcommand{\sn}{\sin}
\newcommand{\cs}{\cos}

\newcommand{\snb}{\sin \beta}
\newcommand{\csb}{\cos \beta}

\newcommand{\lt}{\left}
\newcommand{\rt}{\right}

\newcommand{\lam}{\eta}
\newcommand{\ov}{\over}

\begin{document}
\begin{titlepage}
\begin{flushright}
 KEK Preprint 97-96\\
 KEK-TH- 528\\
 OU-HET 271\\
\today
\end{flushright}
\setcounter{page}{0}
\vspace{2 mm}
\begin{center}
{\Large\bf Non-Decoupling Effects  of Higgs Bosons
        on  $e^+e^- \rightarrow W^+_LW^-_L$ in 
       the Two-Doublet Model}
\end{center} 
\vspace{10 mm}
\begin{center}
{\large \bf Shinya Kanemura 
  \footnote{e-mail: kanemu@theory.kek.jp}}\\    
\vspace{2mm}
{\em Theory Group, KEK,\\
       Tsukuba, Ibaraki 305, Japan}\\
and  \\
\end{center}
\begin{center}
{\large \bf Hide-Aki Tohyama
\footnote{e-mail: tohyama@phys.wani.osaka-u.ac.jp}}\\
\vspace{2mm}
{\em Department of Physics, Osaka University,\\
              Toyonaka, Osaka 560, Japan}\\
\end{center}
\vspace{6 mm}
\begin{abstract}
The non-decoupling effects of heavy Higgs bosons on  the process  
$e^+e^- \rightarrow W^+_LW^-_L$ are discussed in the two Higgs-doublet model. 
The one-loop corrections to the cross section are calculated by using 
the equivalence theorem and the explicit expressions of its deviation 
from the standard model are derived. 
The leading mass contributions to the deviation in expansion by 
$s$ are related to the $T$ parameter by the low-energy theorem. 
The next-to leading ones are free from the present data, which should 
be determined by future experiments. 
The deviation can amount to $\sim 3$ \% at $\sqrt{s} = 1$ TeV
under the constraint from the present data,  
so that it may give useful information on the Higgs sector 
in cooperation with data from future linear colliders.
\end{abstract}
\vspace{22mm}
\end{titlepage}

\section{Introduction}
\hspace*{18pt}      
Radiative corrections can be one of the powerful tools for exploration 
of new physics in cooperation with the experimental precision data. 
The study of the oblique corrections, 
which are controlled by three parameters $S$, $T$ and $U$ \cite{pt}, 
has played a crucial role to determine properties of new physics 
with the data from LEP (and SLC) precision measurement \cite{lep}. 
Standard Model (SM) has not been found to have
any substantial deviation between its prediction and the data,  
while some types of new physics based on the dynamical breaking of 
electro-weak gauge symmetry have been strongly constrained by the data 
by virtue of their non-decoupling properties \cite{hmhk}. 
Generally speaking, radiative corrections can be useful to 
constrain the new physics in which heavy mass effects do not decouple 
from a low energy observable. 

There are models in which non-decoupling effects are expected    
but also are not completely constrained yet by the present data.
To obtain substantial information about such models indirectly, 
we need to study radiative corrections to other types of observable 
than oblique ones, which are expected to be measured at LEP 2 \cite{lep2} 
or future experiments \cite{lc,appi}. 
One of these models may be an extension of SM with two Higgs doublets, 
in which the non-decoupling effects of heavy Higgs bosons may be 
expected to appear as well as those of the fermion contributions.
The two Higgs-doublet model (THDM) has been studied \cite{hhg}
so far with a lot of motivations such as additional CP phases \cite{cp}, 
which are 
often required for the electro-weak baryogenesis \cite{ckn}, 
the strong CP problem and Minimal Supersymmetric Standard Model (MSSM). 
We note that in MSSM, the heavier Higgs bosons ($\mh$, $\mG$ and $\mA$)
can become heavy only by making the soft-breaking parameter to be large 
because all the Higgs self-couplings are fixed to be ${\cal O}(\mw/v^2)$.  
As a result, very heavy Higgs bosons, as well as all heavy super-particles,  
are decoupled from the low energy theory \cite{dec,lis} 
and the non-decoupling effects 
of these particles can be no longer expected in MSSM.    
We still stress that in the non-suspersymmetric THDM, 
there remains a parameter region in which the model becomes a non-decoupling 
theory for heavy Higgs bosons \cite{kt}.

One of the observables for which we can expect to have precision data 
next may be the scattering process $e^+e^- \rightarrow W^+ W^-$.  
This process is not only one of the main target processes at LEP 2 
but also expected to be well measured at future $e^+e^-$ linear colliders.  
Ahn {\it et al.} \cite{apls}
once showed that the non-decoupling effects of heavy fermions 
on this process can substantially be enhanced by the 
factor $s/\mw$ in a region $\mw \ll \sqrt{s} < M$, 
where $M$ represents the heavy fermion mass. 
This enhancement disappears in the high energy limit, $M \ll \sqrt{s}$,  
because of the unitarity cancellation between $s$- and $t$-channel diagrams. 
The heavy mass effects on this process appear 
through the corrections to the triple gauge vertices (TGV's)
$W^+W^-Z^0$ and $W^+W^-\gamma$ \cite{hhpz},   
as well as to the propagators of intermediate $Z^0$ and $\gamma$. 
Appelquist and Wu have studied the connection between TGV form factors  
and the chiral Lagrangian operators \cite{aw}.    
The one-loop estimation of TGV's has been made by several authors 
in SM \cite{apls,papa0},  
in the model with extra fermions such as might be present  
in technicolor-type theories \cite{aw,iltt}, 
and in MSSM \cite{tgv2}.  
The non-decoupling effects of heavy fermions 
have been one of their main interests.
The possibility of the extra fermion generation and 
technicolor-like models, however, 
has already been constrained to considerable extent by careful 
comparison between the oblique parameters and 
the present precision data \cite{hmhk}.  
On the other hand, the non-decoupling effects of 
additional heavy Higgs bosons have not been so studied so far.       
In spite of the constraint from the present data \cite{gr}, 
THDM remains to have large allowed region for the Higgs boson 
masses and the mixing angles.  
Thus it is quite interesting to investigate the possibility to constrain 
the parameter region of THDM through the correction to 
the observables which are to be measured at future measurements. 
This is just our motivation to this work.

In this paper, we discuss the non-decoupling effects of heavy Higgs 
boson masses on the scattering of the polarized electron-positron into a 
longitudinal $W$ boson pair in THDM with a softly broken discrete symmetry.
We investigate the cross section in the one-loop level 
($O(1/(4\pi v)^2)$) and also estimate its deviation from SM, 
$\delta (s) = \sigma_{THDM}/\sigma_{SM}-1$. 
Note that this quantity is related to the deviation  
of the TGV form factors $\kappa_V(s)$ and $g_1^V(s)$,  
which are defined in Ref \cite{hhpz}.  
In calculation, we make full use of the equivalence theorem \cite{et0,et} 
in the Landau gauge and drop the gauge boson loops because of 
the suppression by $m_W^2/M_{Higgs}^2$.
This procedure is quite justified for our purpose to extract 
the non-decoupling effects of heavy Higgs bosons \cite{kt,mw}. 
We obtain the explicit expressions for the leading (quadratic) 
and the next-to leading (logarithmic) contributions 
of the heavy Higgs masses to $\delta(s)$ in expansion by $s/M_{Higgs}^2$.

We find that the leading contributions to $\delta(s)$ 
are written in terms of $\Delta \rho ( =\alpha_f T)$ parameter. 
This is regarded as a kind of the low energy theorem 
(Here low energy means $\sqrt{s} \ll M_{Higgs}$.). 
It is understood from the fact that the physical quadratic contributions 
of the heavy masses comes from an unique term 
$\sim \beta_1 f^2 ({\rm tr}T V_\mu)^2$ in the chiral Lagrangian \cite{aw}.
This term disappears in the custodial $SU(2)_V$ symmetric limit \cite{sc}.
The similar phenomenon has been known to appear in the scattering such as 
$W_L^+ W_L^- \rightarrow W^+_L W_L^-$ \cite{cgg}.  
On the other hand, the next-to leading contributions include 
new additional parameters other than $S$, $T$ and $U$.  
Thus these contributions can give substantial information about the 
new physics. 
In the MSSM like cases, the both contributions turn out to vanish 
in the heavy Higgs limit, so that the heavy mass effects 
are decoupled from the low-energy observable consistently.     
The numerical study shows that the deviation $\delta (s)$ can amount 
to $\sim$ 2-3 \% at $\sqrt{s} = 1$ TeV by the non-decoupling effects
of the Higgs boson masses even under the constraint 
from the present data as well as from the perturbative unitarity \cite{kkt}.
We note that the accuracy may be expected to be smaller than $\sim$ 2 \% 
at $\sqrt{s} \sim 1$ TeV by taking account of the ambiguity due to 
our approximation as well as the ambiguity of the measurement by the 
expected statistic and systematic errors at future $e^+e^-$ linear colliders 
\cite{tgv}. 
Therefore we conclude that such deviation between THDM and SM on the process 
may be detectable at future experiments and bring useful information 
on the Higgs sector.
 
In Sec.2, we will summarize the results by the effective Lagrangian briefly. 
THDM will be defined in Sec.3. Details of calculation will be shown in Sec.4. 
The concrete expressions for the deviation from SM of the cross section are 
derived and the non-decoupling properties are discussed in Sec.5. 
Results will be summarized in the last section.

\section{$W^+W^-Z^0$,  $W^+W^-\gamma$ vertices}
\hspace*{18pt}The contribution of non-decoupling effects to 
$e^+e^- \rightarrow W^+W^-$ comes from  
the corrections to the triple gauge vertices (TGV's) \cite{hhpz} 
as well as the oblique-type ones. 
All the corrections in other types of diagram 
are suppressed by the electron masses or gauge boson masses. 
The effective Lagrangian for TGV's in the $C$, $P$ and $CP$ conserving case 
is expressed as     
\begin{eqnarray}
  \fr{{\cal L}^{WWV}}{g_{WWV}}
&=& i g_1^V (W_{\mu\nu}^+W^{-\mu}V^{\nu} - W_{\mu\nu}^-W^{+\mu}V^{\nu})
    + i \kappa_V W_\mu^+ W_{\nu}^-V^{\mu\nu} \nn \\
& & + i\fr{\lambda_V}{(4 \pi v)^2} W_{\mu\nu}^+
      W^{-\nu}_{\;\;\;\;\;\rho} V^{\rho\mu}, 
\end{eqnarray}
where $g_{WW\gamma} = -e$ and $g_{WWZ} = - e \cot \theta_W$.
The tree level form factors are given in SM and also in THDM as 
\begin{eqnarray}
  g_1^V = \kappa_V = 1, \;\; \lambda_V = 0.
\end{eqnarray}
The deviation from these tree-level values is generated at the loop level 
and they are denoted here as $\Delta g_1^V = g_1^V - 1$ and the same 
for others.    
It is known that $\kappa_\gamma$ and $\lambda_\gamma$ are related to 
the magnetic moment and quadrupole moment of $W^\pm$-bosons \cite{hhpz}.

Appelquist and Wu have derived the relation between the form factors 
and the coefficients in the chiral Lagrangian \cite{aw}. 
From the context of the power counting method,  
it is expected that at one loop level 
there are quadratic mass contributions of inner heavy particles with the mass 
$M$ in $g_1^V$ and $\kappa_V$; namely 
$\sim O(M^2/(4\pi v)^2)$, 
where $v$ is the vacuum expectation value (VEV).  
These quadratic contributions actually 
appear if the new physics does not have the custodial $SU(2)_V$ invariance 
\cite{sc}.   These occur through the dimension 2 operator \cite{aw}
\begin{eqnarray}
  {\cal L}_1' \equiv \fr{1}{4} \beta_1 (4 \pi v)^2 
         \lt[ {\rm tr}(T V_\mu) \rt]^2,  \label{L1}
\end{eqnarray}
where $V_\mu$ and $T$ are 
expressed in terms of the dimensionless unitary unimodular matrix field 
$U(x)$ as $V_\mu = (D_\mu U)U^\dagger$ and $T = U \tau_3 U^\dagger$.
The dimensionless parameter $\beta_1$, which measures 
the breaking of $SU(2)_V$, is known to be related  
to $\alpha_f T$ $(= \Delta \rho)$ as $\al_f T = 2 \be_1$, where 
$\alpha_f$ is the fine structure constant.    

Note that the form factors, $g_1^V$, $\kappa_V$ and $\lambda_V$,  
should be considered as functions of the energy $\sqrt{s}$ in general.   
The next-to leading mass contribution 
in expansion by $s$ to the form factors may become 
important at high energy region \cite{papa0}.  
Since these contributions have not been known yet, 
this can be expected to bring additional information for new physics 
in cooperation with the measurement at future linear colliders \cite{lc}.  
The helicity amplitudes for the polarized electron-positron 
scattering into a $W$-boson pair are expressed in terms 
of these form factors \cite{hhpz}. 
In the case of longitudinally polarized $W$-boson final states, 
the helicity amplitudes depend only on the combination of  
$g_1^V(s) + (s/2\mw) \kappa_V(s)$ 
even if there is no $C$, $P$, or $CP$ invariance 
in the model, where $V = \gamma$ and $Z^0$.     
We evaluate the correspondence to these form factors later.

\section{Two Higgs-Doublet Model}
\hspace*{18pt}
Here we define THDM with a softly broken discrete symmetry; 
$\Phi_1 \rightarrow \Phi_1$, $\Phi_2 \rightarrow -\Phi_2$.
This model is the most general one 
in which the natural flavor conservation is realized \cite{gw}.
Since the effects of CP violation disappear 
in the process $e^+e^- \rightarrow W^+_LW^-_L$ \cite{hhpz} as already 
mentioned in Sec. 2, 
we consider the CP invariant Higgs sector from the beginning. 
The Lagrangian of the Higgs sector is then given as 
\begin{eqnarray}
 {\cal L}(\Phi_1,\Phi_2) &=&
  \mu_1^2\mid \Phi_1 \mid^2
  +   \mu_2^2\mid \Phi_2 \mid^2                 
 + 2 \mu_3^2   {\rm Re} \Phi_1^\dagger \Phi_2        \nonumber \\
 &&- \lam_1 \mid \Phi_1 \mid^4
     -\lam_2 \mid \Phi_2 \mid^4           
     -\lam_3 \mid \Phi_1 \mid^2 
      \mid \Phi_2 \mid^2 \nonumber \\
 &&- \lam_4({\rm Re} \Phi_1^\dagger \Phi_2)^2  
  - \lam_5({\rm Im} \Phi_1^\dagger \Phi_2)^2.          \label{lag}
\end{eqnarray}
The Higgs sector then includes the eight parameters in general. 
We here consider all the parameters to be free.  
( Note that MSSM is considered as a special case in this 
  Lagrangian, in which the supersymmetry imposes strong relations 
  between these parameters. 
  In MSSM, all the quartic couplings are constrained into $O(g^2)$, where $g$ 
  denotes weak gauge couplings. 
  Thus in the heavy Higgs limit, which 
  is realized by $\mu_3^2 \rightarrow \infty$, 
  the Higgs mass-effects are decoupled from low-energy observables 
  \cite{dec,lis}. )

The Higgs doublets, both of which are assigned hypercharge as $Y=1/2$, 
are parametrized as 
\begin{eqnarray}
\Phi_{i}&=& \left(\matrix{  w_{i}^+ \cr
       { 1 \over \sqrt{2}}
      (v_{i} + h_{i} + i z_{i}) \cr}\right),\; (i = 1,2)
\end{eqnarray}
where vacuum expectation values $v_1$ and $v_2$ 
satisfy $\sqrt{v_1^2 + v_2^2} = v \sim 246$  GeV.
The mass eigenstates are obtained by rotating the 
fields in the following way:
\begin{eqnarray}
\left( \matrix{
h_{1}^0 \cr
h_{2}^0 \cr }\right)&=&\left( \matrix{
\cos\alpha & - \sin\alpha \cr
\sin\alpha &   \cos\alpha \cr}\right)
\left( \matrix{
H^0 \cr
h^0 \cr}\right), \\
\left( \matrix{
w_{1}^\pm \cr
w_{2}^\pm \cr}\right)&=&\left( \matrix{
\cos\beta & -\sin\beta \cr
\sin\beta &  \cos\beta \cr}\right)
\left( \matrix{
w^\pm \cr
H^\pm \cr}\right), \\
\left( \matrix{
z_{1}^0 \cr
z_{2}^0 \cr}\right)
&=&\left( \matrix{
\cos\beta & -\sin \beta \cr
\sin\beta & \cos \beta \cr}\right)
\left( \matrix{
z^0 \cr
A^0 \cr}\right).
\end{eqnarray}
By setting $\tan \be = v_2/v_1$, $w^\pm$ and $z^0$ become 
the Nambu-Goldstone bosons which are to be absorbed into the  
longitudinally polarized gauge bosons $W^\pm_L$ and $Z^0_L$, respectively.
$H^\pm$ and $A^0$ are then massive charged and CP-odd neutral states. 
On the other hand, $h^0$ and $H^0$ are massive CP-even neutral states. 
The mixing angle $\al$ is taken in order $h^0$ to be lighter than $H^0$.

The quartic couplings are then represented by using 
the mass parameters, the mixing angles and VEV as  
\begin{eqnarray}
\lam_1 &=& {1 \ov 2 v^2 \cos^2 \be}
           (\mh \cos^2 \al +\mH \sin^2 \al) -
           {\mu_3^2 \ov 2 v^2} {\snb \ov \cos^3 \be} \\
\lam_2 &=& {1 \ov 2 v^2 \sin^2 \be}
           (\mh \sin^2 \al +\mH \cos^2 \al) -
           {\mu_3^2 \ov 2 v^2} {\csb \ov \sin^3 \be}      \\
\lam_3 &=& {\sin 2 \al \ov v^2 \sin 2 \be}(\mh -\mH) +
            {2 \mG \ov v^2} -{2 \mu_3^2 \ov v^2 \sin 2 \be}   \\
\lam_4 &=& -{2 \mG \ov v^2}+  {4 \mu_3^2 \ov v^2 \sin 2 \be} \\
\lam_5 &=& {2 \ov v^2}(\mA-\mG)     
\end{eqnarray}
In general, THDM does not have the custodial $SU(2)_V$ symmetry.  
But if $\lam_5$ is zero, the Higgs sector turns out to be  
$SU(2)_V$ symmetric even after the spontaneous breakdown of 
$SU(2)_L \otimes U(1)_Y$ 
gauge invariance occurs \cite{haber}. 
Therefore the mass splitting between 
$H^\pm$ and $A^0$ measures $SU(2)_V$-breaking in the Higgs sector 
\cite{kt,haber}.
Note that there are some cases where we can take the mass splitting to be 
enough large within the constraint from the present data. 
For example, if we set $\al - \be = \pi/2$ and $\mG \sim m_{H^0}^2$, 
$\mA$ and $m_{h^0}^2$ can be chosen freely with keeping $\al_f T \sim 0$.    

As for the Yukawa couplings, there can be two types of model
in THDM (what we call, Model I and II in Ref. \cite{hhg})  
in which natural flavor conservation is realized 
by imposing discrete symmetries (see eq. \eq{lag}). 
Note that the difference between Model I and II vanishes 
in such the situation as we will consider later,
where the mass of bottom quarks is negligible.

\section{$e^+ e^- \rightarrow W^+_L W^-_L$ in THDM}
\hspace*{18pt}      
In this section, we show the one-loop calculation of the
process $e^-_X e^+_Y \rightarrow W^+_L W^-_L$ in THDM, 
where $X$ ($Y$) is the helicity of the electron (positron).        
The calculation is quite simplified by making full use of 
the equivalence theorem (ET) \cite{et0}, which says that 
in the case of $\sqrt{s} \gg m_W$  
the cross section for $e^+ e^- \rightarrow W^+_L W^-_L$ is 
equivalent to that for $e^+ e^- \rightarrow w^+ w^-$ 
up to ${\cal O}(m_W^2/s)$ 
\footnote{
  In general, the error is of ${\cal O}(m_W/\sqrt{s})$. 
  It turns out to be ${\cal O}(m_W^2/s)$ in some cases 
  including present process \cite{ps}.}. 
The extension of ET to the loop level 
has been studied by several authors \cite{et}. 
They found that the some modification factors,  which depend 
on gauge parameters, should be multiplied.    
He {\it et al.} showed that in SM the Landau gauge is a good choice 
for such the purpose here because the modification 
factors no longer depend on the Higgs boson mass and 
they can be set into unity within the approximation.   
We note that this situation is not changed even in the case of THDM 
\cite{kt,kt0}. 
Thus the radiative correction here is calculated in the Landau gauge.     

The non-decoupling effects of Higgs boson masses  
on  $e^+ e^- \rightarrow w^+ w^-$    
come from the corrections to 
the $V^\mu w^+w^-$ vertices, $(V = \gamma,\; Z^0)$ in 
the $s$-channel gauge boson intermediate diagrams.  
Other types of diagrams    
(the neutrino exchanged $t$-channel and box-type diagrams)  
are always suppressed by powers of the electron mass or gauge boson masses.  
The oblique corrections in the $s$-channel diagrams are also neglected 
because of the suppression factor $\mw/M_{Higgs}^2$. 
(Note that this approximation is valid in the situation like 
 $\mw \ll s < M^2_{Higgs}$ or $\mw \ll M^2_{Higgs} < s$.)
Thus we have only to calculate corrections to the 
$V^\mu w^+w^-$ vertices for our purpose here. 
Moreover, it turns out that 
only the Higgs-Goldstone boson loops contribute to the vertices 
because the diagrams including a gauge boson loop are 
relatively suppressed by a factor $m_W^2/M_{Higgs}^2$ \cite{kt,mw}.       

The  $V^\mu w^+w^-$ vertices can be decomposed as
\begin{eqnarray}
  i {\cal M}_{Vww}^\mu(s) = i {\cal M}_{Vww}^-(s) (p_+ - p_-)^\mu
                      + i {\cal M}_{Vww}^+(s) (p_+ + p_-)^\mu,    
\end{eqnarray}
where $p_+$ ($p_-$) is the momentum of $w^+$ ($w^-$).     
Since the second term of RHS, which is proportional to 
$(p_+ + p_-)^\mu$, produces the contribution suppressed by 
the negligible electron mass squared, we have only to 
calculate ${\cal M}_{Vww}^-(s)$ for our purpose.
It is expressed by      
\begin{eqnarray}
  i {\cal M}_{Vww}^-(s) = - i g_V \;
\lt\{ 1+{\cal G}_{V}(s)+ Z_w + \fr{\delta g_{V}}{g_{V}}\rt\} 
\equiv -i g_V \Gamma_V(s), \label{12}
\end{eqnarray}
where ${\cal G}_{V}(s)$ are the contributions of one-loop diagrams 
other than counterterms,  
$Z_w$ denotes the wave function renormalization constant for the external 
$w^{\pm}$ lines and $\delta g_{V}$ are the shift of coupling constants defined
by $g_V \rightarrow g_V + \delta g_V$. 
The tree level coupling constants $g_V$ are given by
\begin{eqnarray}
  g_{Z}       =    e \cot 2 \theta_W,\;\;\;\;  g_{\gamma}  =    e, 
\end{eqnarray}
where $e$ and $\theta_W$ denote the electric charge and
the Weinberg angle respectively.

The scattering amplitude for the polarized $e^+e^-$ scattering 
into $w^+ w^-$ is given in terms of $\Gamma_V(s)$ as 
\begin{eqnarray}
i {\cal A}(e^-_X e^+_Y \rightarrow w^+ w^-) 
= i e^2 \bar v_X \gamma_{\mu} u_X    
  \left\{{\Gamma_{\gamma}(s) \over s}+ f_{XY}
{\Gamma_{Z}(s) \ov s-\mz}  \right\}(p_+ - p_-)^{\mu}, 
\end{eqnarray}
where $f_{XY}$ is defined according to the $e^-e^+$ helicities by 
\begin{eqnarray}
  f_{LR} =   \cot 2 \theta_W, \;\; 
  f_{RL} =  - \fr{\cos 2 \theta_W}{2 \cos^2 \theta_W}.   \nn
\end{eqnarray}
The total cross section is then expressed in each case as 
\begin{eqnarray}
\sigma (e^-_X e^+_Y \rightarrow w^+ w^-) =   
{ e^4 s \ov 24 \pi}  
 \lt|  {\Gamma_{\gamma}(s) \over s}+ f_{XY}
       {\Gamma_{Z}(s) \ov s-\mz}   
 \rt|^{\;2} \!\!\!.  \label{cross}
\end{eqnarray}

The estimation of $\Gamma_V(s)$ is performed in order. 
Here we concentrate into the contributions from the Higgs sector.  
At first, the one-loop contributions to ${\cal G}_V(s)$ are calculated as
\begin{eqnarray}
{\cal G}_{Z}(s) &=& {1 \ov (4 \pi v)^2}
\left[ {2 \ov \cos 2 \theta_W}
       \left\{ {\tilde C}[A^0 H^{\pm} H^0] \sin^2 \ab   \rt.\rt. \nn\\
  &&   \;\;\;\;\;\;\;\;\;\;\;\; \;\;\;\;  \;\;\;\;\;\;\;\;\;\;\;\;   
    +   \lt. {\tilde C}[A^0 H^{\pm} h^0] \cos^2 \ab  \rt\} \nn \\  
&+& 
{\tilde C}[w^\pm H^0 w^\pm] \cos^2 \ab 
+ {\tilde C}[w^\pm h^0 w^\pm] \sin^2 \ab \nn  \\   
&+& 
{\tilde C}[H^{\pm} H^0 H^{\pm}] \sin^2 \ab +  
{\tilde C}[H^{\pm} h^0 H^{\pm}] \cos^2 \ab  \nn   \\ 
&+& \left. {\tilde C}[H^{\pm}A^0 H^{\pm}] 
    \begin{array}{c} \; \\ 
                  \; 
     \end{array} \!\!\!\! \right],
\end{eqnarray}
and
\begin{eqnarray}
{\cal G}_{\gamma}(s) &=& {1 \ov (4 \pi v)^2} 
\left\{
{\tilde C}[w^\pm H^0 w^\pm] \cos^2 \ab +
{\tilde C}[w^\pm h^0 w^\pm] \sin^2 \ab      \right.   \nn\\
&+&  
{\tilde C}[H^{\pm} H^0 H^{\pm}] \sin^2 \ab +  
{\tilde C}[H^{\pm} h^0 H^{\pm}] \cos^2 \ab    \nn   \\
&+&  \left.  
{\tilde C}[H^{\pm} A^0 H^{\pm}]     \right\},         
\end{eqnarray}
where $p_+^2=p_-^2=0$ is taken in the Landau gauge.
The function ${\tilde C}[123]$ is defined by
\begin{eqnarray}
{\tilde C}[123] &=&  (m_1^2-m_2^2)(m_3^2-m_2^2)(C_{11}-C_{12})[123],
\end{eqnarray}
where  
$C_{11}[123]$ and $C_{12}[123]$ can be written 
in terms of the $B_0$ and $C_0$ functions introduced by 
Passarino and Veltman \cite{pv}. 
We employ these notations here according to the definition in 
Ref.\cite{hmhk}.  
Secondly, the wavefunction renormalization $Z_w$ is calculated 
up to ${\cal O}(\mw/M_{Higgs}^2)$ as \cite{kt,kt0}
\begin{eqnarray}
  Z_w &=& -{1 \ov (4 \pi v)^2} \lt\{ {1 \ov 2}(2 \mG +\mH +\mh +\mA) \rt. \nn\\
      &&+ {\mG \mA \ov \mA - \mG} \ln {\mG \ov \mA}
       +  \cos^2 \ab {\mG \mH \ov \mH - \mG} \ln {\mG \ov \mH} \nn \\
      &&+ \lt.\sin^2 \ab {\mG \mh \ov \mh - \mG} \ln {\mG \ov \mh}  \rt\}
\end{eqnarray}
Thirdly, 
the renormalization of the coupling constants $\delta g_V$ is expressed as
\begin{eqnarray}
\fr{\delta g_{Z}}{g_{Z}}       &=& 
{1 \ov 2 \cos^2 2 \theta_W}
\left ({\delta m_Z^2 \ov m_Z^2} - 
       {\delta m_W^2 \ov m_W^2} -
       \sin^2 2 \theta_W {\delta \al_f \ov \al_f} \right),  \label{gz} \\
\fr{\delta g_{\gamma}}{g_{\gamma}}  &=& \fr{1}{2} {\delta \al_f \ov \al_f},
\end{eqnarray}
where $\delta \al_f$ is the shift of the fine structure constant and  
$\delta m^2_V$ are mass renormalizations defined as 
$\al_f \rightarrow \al_f + \delta \al_f$ and 
$m_V^2 \rightarrow m_V^2 + \delta m_V^2$ (here $V = W$ or $Z$), respectively. 
Note that $\delta \al_f$ is relatively suppressed by the factor 
$\mw/M^2_{Higgs}$. 
On the other hand, $\delta m^2_V$ are expressed up to 
${\cal O}(\mw/M^2_{Higgs})$ as \cite{tou}
\begin{eqnarray}
{\delta m_W^2 \ov m_W^2} 
&=&
 {- 1 \ov (4 \pi v)^2} \lt\{ {1 \ov 2}(2 \mG +\mH +\mh +\mA) 
+ {\mG \mA \ov \mA - \mG} \ln {\mG \ov \mA}  \rt.      \nn   \\
&&      \!\!\!\!\!\!\!\!\!\!\!\!\!\!\!\!\!\!
+ \lt.  \cos^2 \ab {\mG \mH \ov \mH - \mG} \ln {\mG \ov \mH} 
+ \sin^2 \ab {\mG \mh \ov \mh - \mG} \ln {\mG \ov \mh}  \rt\},\nn \\
 {\delta m_Z^2 \ov m_Z^2} 
&=& { - 1 \ov (4 \pi v)^2} \lt\{ {1 \ov 2}(\mH +\mh +\mA) 
+  \cos^2 \ab {\mA \mH \ov \mH - \mA} \ln {\mA \ov \mH}\rt. \nn \\
&&+ \lt. \sin^2 \ab {\mA \mh \ov \mh - \mA} \ln {\mA \ov \mh}  \rt\}   . 
\end{eqnarray}

By inserting all these results into eq. \eq{cross}, 
we finish to calculate the total cross sections 
at one loop level.

\section{Non-Decoupling Effects of Higgs Bosons}
\hspace*{18pt}      
Now we consider the non-decoupling effects of heavy Higgs masses 
on the ratio $R(s)=\sigma_{THDM}/\sigma_{SM}$, where the cross section 
in SM, $\sigma_{SM}(s)$, is to be calculated in the one-loop level 
in the same approximation manner as $\sigma_{THDM}(s)$.   
The one-loop corrections to $\Gamma_V$ (see eq. \eq{12}) in SM 
case can be then immediately calculated as  
\begin{eqnarray}
  \Gamma_{\gamma, Z}^{SM} 
    = 1 + \fr{1}{(4\pi v)^2} \lt\{ 
       \tilde{C}[w^\pm\phi^0_{SM}w^\pm ] 
      -  \fr{1}{2} m_{\phi^0_{SM}}^2  \rt\},
\end{eqnarray}
where $\phi_{SM}^0$ is the Higgs boson in SM. 
For the case of $e^-e^+$ helicity $LR$ ($RL$),
the magnitude of $\sigma_{SM}(s)$ amounts at one loop level  
to 355 $\sim$ 358 (67.6 $\sim$ 68.3) fb at $\sqrt{s} = 500$ GeV and 
98.9 $\sim$ 98.5  (19.7 $\sim$ 19.6) fb at $\sqrt{s} = 1000$ GeV 
for $m_{\phi^0_{SM}} = 140 \sim 1000$ GeV, respectively.   
These values are consistent with the previous results \cite{apls,tgv} 
up to the ambiguity due to the use of the equivalence theorem. 
We did not include the fermion-loop contribution in these values. 
This is because that it takes the same form between THDM and SM, so that 
it consequently does not contribute to the deviation at all.

It is convenient to parametrize the ratio $R(s)$ as 
\begin{eqnarray}
 R(s) =  \fr{\sigma_{THDM}}{\sigma_{SM}} =  1 + \delta(s).
\end{eqnarray}
The deviation-function $\delta(s)$ is expressed in terms of  
the difference of the one loop corrections to $\Gamma_V$  between 
THDM and  SM; 
\begin{eqnarray}
  \delta(s) =  2 {\rm Re}\lt[ 
\fr{\fr{1}{s} \lt\{ \delta \Gamma_\gamma^{THDM} -
                    \delta \Gamma_\gamma^{SM}     \rt\}
    +  \fr{f_{XY}}{s - m_Z^2} 
              \lt\{ \delta \Gamma_Z^{THDM} -
                         \delta \Gamma_Z^{SM}   \rt\}}
{\fr{1}{s} + \fr{f_{XY}}{s - m_Z^2}} \rt], \label{delta}
\end{eqnarray}
where one-loop corrections $\delta \Gamma_V^{model}$ are defined by
$\Gamma_V^{model} = 1 + \delta\Gamma_V^{model}$. 
The quark loops do not contribute to $\delta (s)$ 
at all because of the exact cancellation between both models. 
As we mentioned before, $\delta (s)$ is expressed in terms of   
$\Delta g_1^V(s)$ and $\Delta \kappa_V(s)$. 
We have the relation for $\mw \ll s$ as 
\begin{eqnarray}
  \delta (s) &=& \fr{4 \gamma^2}{1 + f_{XY}}
     \lt\{ \lt( 1 - \fr{\xi_{XY}}{2 \sin^2 \theta_W }\rt) 
      \lt(\Delta \kappa_Z^{THDM}(s) - \Delta \kappa_Z^{SM}(s)\rt) \rt. \nn \\
 &&  \lt.  \begin{array}{c} \; \\ 
                  \; 
     \end{array}\;\;\;\;\;\;\;\;\;\;\;\;\;\;  -
 \lt(\Delta \kappa_\gamma^{THDM}(s) - \Delta \kappa_\gamma^{SM}(s) \rt) \rt\},
\end{eqnarray}
where $\xi_{LR}= 1$, $\xi_{RL} = 0$ and $\gamma = \sqrt{s}/2m_W$.

For extracting non-decoupling effects of heavy Higgs bosons, 
we expand $\delta(s)$ by powers of $s$ as 
\begin{eqnarray}
  \delta(s) = \delta^{(0)} + 
      \delta^{(1)} s + O (s^2).
\end{eqnarray}
Note that this expansion is valid 
only in the case with  $m_W \ll \sqrt{s} \ll M_{Higgs}$. 
At one loop level ${\cal O}(1/(4 \pi v)^2)$, 
all the non-decoupling effects are included in 
$\delta^{(0)}$ and $\delta^{(1)}$. 
By dimensional counting, we know that 
$\delta^{(0)}$ represents the quadratic Higgs 
mass effects and $\delta^{(1)}$ includes at most logarithmic ones. 

We show the explicit expressions for  
$\delta^{(0)}$ and $\delta^{(1)}$ at one loop level.
In calculation, we identify the lighter neutral Higgs boson $h^0$ in THDM 
as SM like Higgs boson $\phi^0_{SM}$.  
At first,  $\delta^{(0)}$ is calculated as
\begin{eqnarray}
  \delta^{(0)}&=& \fr{1}{(4\pi v)^2} 
                  \fr{f_{XY}}{1 + f_{XY}}
     \lt( \fr{2}{\cs 2\theta_W} + 
                 \fr{1}{\cs^2 2\theta_W} \rt) \nn \\
  &\times& \lt[ F (\mG,\mA)\rt.  \nn \\
&&     + \sn^2 \ab \lt\{   F (\mG,\mh) 
                        - F (\mA,\mh) \rt\}  \nn \\
&& \lt. + \cs^2 \ab \lt\{  F (\mG,\mH)
                        - F (\mA,\mH)  \rt\} \rt],  
\label{d0}
\end{eqnarray}
where 
\begin{eqnarray}
F(x,y) =  \fr{x + y}{2} 
             - \fr{x y} {x - y}\ln \fr{x}{y}.   
\end{eqnarray}
Comparing eq. \eq{d0} to the expression of the $\Delta \rho$ $(= \al_f T)$ 
parameter \cite{hhg}, 
we have a relation
\begin{eqnarray}
   \delta^{(0)} = \fr{f_{XY}}{1 + f_{XY}} \lt( \fr{2}{\cs 2\theta_W} + 
                 \fr{1}{\cs^2 2\theta_W} \rt) \al_f T.
\end{eqnarray}
The second term in the bracket of RHS comes from $\delta g_Z$ in eq. \eq{gz}. 
We can see that the leading effects, $\delta^{(0)}$, 
can be written in terms of $T$.  
This phenomenon is due to the low energy theorem 
and is understood as the concrete realizations of the fact 
which we discussed in Sec 2.  
In fact, this leading contribution $\delta^{(0)}$ 
vanishes if mass degeneracy between $A^0$ and $H^\pm$ exists. 
In this case, the Higgs sector becomes custodial $SU(2)_V$ symmetric, 
so that the term \eq{L1} is then forbidden.    
We note all the contributions to eq. \eq{d0} come from only 
$\delta \Gamma_Z$.    
As a result, the leading contribution is found not to be substantial because 
$T$ has already been fairly constrained by the present data.

The next-to leading contributions, $\delta^{(1)}$,    
may be possible candidates for the probe of the new physics.
They include non-decoupling effects like $\sim \log M_{\rm Higgs}$.
These are extracted from eq. \eq{delta}  as 
\begin{eqnarray}
  \delta^{(1)}&=& 
  \fr{2}{(4\pi v)^2} 
  \lt[ \lt\{ G(0,\mh,0) - G(0,\mH,0) \rt\} 
            \cos^2 \ab \rt.   \nn\\
&&+  
G(\mG,\mh,\mG) \sin^2 \ab +  
G(\mG,\mH,\mG) \cos^2 \ab    \nn   \\
&&+  \left.  
  G(\mG,\mA,\mG)     \right] / (1 + f_{XY}) \nn \\
&&+ \fr{2 f_{XY}}{(4 \pi v)^2}
\lt[
\begin{array}{c} \; \\ \; 
\end{array} \!\!\!\! 
\lt\{G (0,\mh,0) - G (0,\mH,0) \rt\} \cs^2 \ab \rt.
      \nn  \\ 
 &&+{2 \ov \cos 2 \theta_W}
       \left\{ G (\mA,\mG,\mh) \sin^2 \ab   
    +    G (\mA,\mG,\mH) \cos^2 \ab  \rt\} \nn \\   
&&+ 
G (\mG, \mh, \mG) \sin^2 \ab +  
G (\mG, \mH, \mG) \cos^2 \ab  \nn   \\ 
&&+ \lt. G(\mG,\mA,\mG) 
\begin{array}{c} \; \\ \; 
\end{array} \!\!\!\! \right] / (1 + f_{XY}), \label{d1}
\end{eqnarray}
where $G(x,y,z)$ is the coefficient of the second term of $s$-expansion 
for $\tilde{C}$ function.   This is expressed as 
\begin{eqnarray}
  G(x,y,z) = \fr{1}{6(x-z)} 
           \lt\{ 2 g_0(x,y,z)  
                 +  g_1(x,y,z) + g_1(y,z,x) + g_1(z,x,y) \rt\}, 
\label{log2} 
\end{eqnarray}
where the functions $g_0(x,y,z)$ and $g_1(x,y,z)$ are defined by 
\begin{eqnarray}
  g_0(x,y,z) &=& \fr{- 1}{(x - y)(y - z)(z - x)}
     \lt\{ x^2 z^2 - x y z (x + z) + y^2 (x^2 - x z + z^2) \rt\}, \nn \\ 
\\          
  g_1(x,y,z) &=& \fr{- 1}{(x - z)^2(x - y)^2}
     x^2(y - z) \lt\{ x^2 + x(y + z) - 3 y z \rt\}\ln x. 
\end{eqnarray}
\begin{figure}[t]
\centering{
\leavevmode
\psfig{file=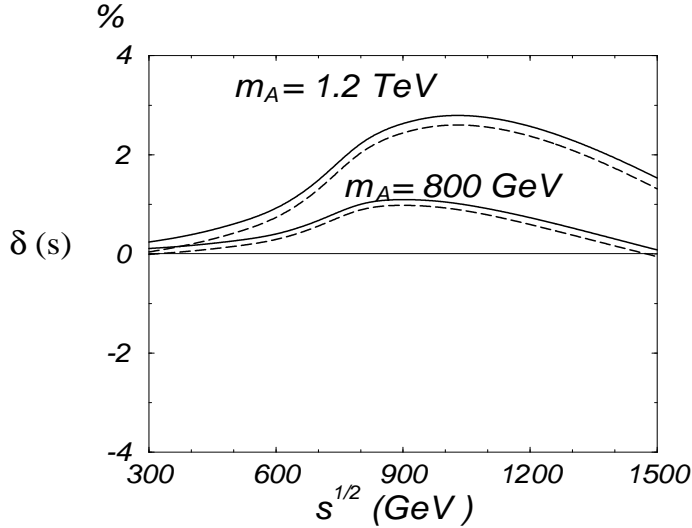,height=70mm,width=90mm,angle=-90}
\vspace{5mm}
}
\caption{
 The $\protect{\sqrt{s}}$ dependence of $\delta(s)$ for  
 $\protect{m_{A^0}} =$ 800 and 1200 GeV.  
 The solid (dashed) lines represent $\protect{\delta(s)}$ for the 
 initial helicity states $\protect{e^-_Le^+_R}$ ($\protect{e^-_Re^+_L}$).
 The other parameters are fixed being taken account of the constraint from 
 $T$ parameter by $m_{h^0} = 140$, $m_{H^0}=350$, $m_{H^\pm}= 347$ GeV, 
 and $\alpha - \beta = \pi/2$.}. 
\end{figure}
Note that 
the function $G(x,y,z)$ vanishes if and only if we set $x = y$ or $y = z$.  
We can see from eq. \eq{d1} that $\delta^{(1)}$ does not vanish  
except for a few cases. 
One of the cases where $\delta^{(1)}$ vanishes 
is that with the complete degeneracy between all of the Higgs boson masses.
Another one is the case with $\al - \be \sim \pi/2$ 
and $\mh \sim \mG \sim \mA$. 
We note that the latter case occurs in MSSM with the large $m_{A^0}$ limit 
\cite{hhg}.     
Since these cases also imply the custodial $SU(2)_V$ invariance in 
the Higgs sector,  
we find that the non-decoupling effects both $\delta^{(0)}$ 
and $\delta^{(1)}$ vanish simultaneously in these cases and 
the model then becomes a decoupling theory for Higgs boson masses 
\cite{dec,lis}.  
On the other hand, 
there are some cases in which $\delta^{(1)}$ becomes large to some extent 
with parameters satisfying the constraint from the present experimental data.
For example, if we set $m_{H^0} \sim m_{H^\pm}$ and $\alpha - \beta = \pi/2$, 
the other masses $m_{h^0}$ and $m_{A^0}$ can be chosen freely with 
keeping $\alpha_f T \sim 0$.  
In such cases, we can expect to obtain some useful information 
through this process.  
\begin{figure}[t]
\centering{
\leavevmode
\psfig{file=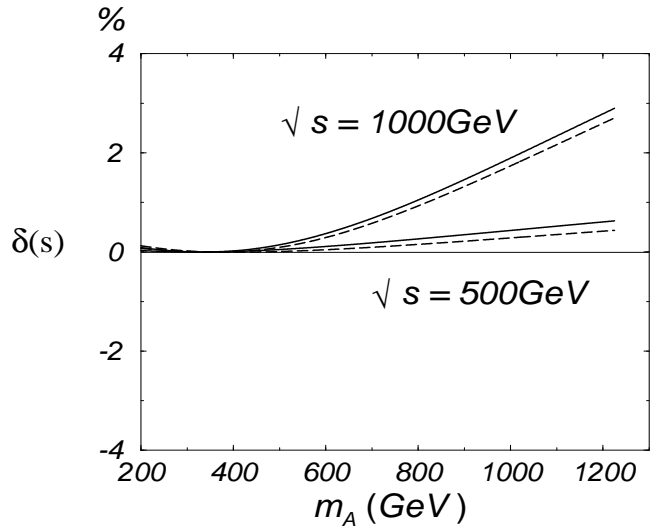,height=70mm,width=90mm,angle=-90}
\vspace{5mm}
}
\caption{
 The $m_A$ dependence of $\delta(s)$ at $\protect{\sqrt{s}} =$ 500 and 
 1000 GeV.  The solid (dashed) lines represent $\protect{\delta(s)}$ for the 
 initial helicity states $\protect{e^-_Le^+_R}$ ($\protect{e^-_Re^+_L}$).
 The other parameters are fixed being taken account of the constraint from 
 $T$ parameter by $m_{h^0}=140$, $m_{H^0}=350$, $m_{H^\pm}= 347$ GeV, 
 and $\alpha - \beta = \pi/2$.}. 
\end{figure}

In the situation such as the momentum $\sqrt{s}$ 
comparable to the largest mass of the Higgs bosons, the question 
whether the non-decoupling effects become relatively 
large or not may occur.  
In this case, the expansion above is no longer allowed, 
so that we have to investigate the non-decoupling effects only
through numerical estimation.
The $\sqrt{s}$ dependence of $\delta(s)$ is described in Figure 1. 
The parameters are chosen in order to satisfy the constraint from the 
present data. 
We here chose $\al - \be = \pi/2$, $m_{h^0} = 140$, $m_{H^0} = 350$, and 
$m_{H^\pm} = 347$ GeV for satisfying 
$- 3.8 \times 10^{-3} < \al_f T_{THDM} < 2.6 \times 10^{-4}$ \cite{hmhk},
where $T_{THDM}$ is the additional contribution to $T$ parameter in THDM.
We can see that the behavior changes
according to the relative energy scale to $M_{Higgs} \sim m_{A^0}$.
The deviation by non-decoupling effects is enhanced by $s$ 
for $\sqrt{s} < m_{A^0}$ 
but is reduced for very high energy region as $m_{A^0} \ll \sqrt{s}$. 
This behavior is consistent with the result for the fermion effects 
by Ahn {\it et al.} \cite{apls} that the enhancement disappears 
in the high energy limit because of the unitarity cancellation 
between $s$- and $t$-channel diagrams. 
In Figure 2, we show  
the $m_A$-dependence of $\delta(s)$ at 
$\sqrt{s} = 500$ GeV and $1000$ GeV in the same choice for other parameters 
as Fig.1. 
We can see in Figs. 1 and 2 that 
the large mass difference (around $\sqrt{s}$) 
between $m_{A^0}$ and $m_{H^0} \sim m_{H^\pm}$ tends to produce 
the large deviation. 
At $\sqrt{s}=1000$ GeV, it amounts to $\sim$ 3 \% for  
$m_{A^0} = 1200$ GeV. 
The deviation for helicity LR is larger than that for RL in general.  
Note that all the parameter choice here is also taken account of 
the constraint from the perturbative unitarity \cite{kkt}.

\section{Conclusion}
\hspace*{18pt}      
We have discussed the non-decoupling effects of the heavy Higgs bosons 
on the scattering process 
$e^-_Le^+_R$ (or $e^-_R e^+_L $) $\rightarrow W^+_LW^-_L$ in THDM.
The cross section has been calculated at one-loop level $O(1/(4\pi v)^2)$ 
by making full use of the equivalence theorem.  
The effects of heavy Higgs bosons have extracted in  
the ratio of the cross section between THDM and SM,
 $R(s)( = 1 + \delta(s))$. 
The leading (quadratic) contributions of the masses 
to $\delta(s)$ become to be written in terms of $T$-parameter.
This phenomenon is regarded as the result by the low energy theorem 
and can be understood by the chiral Lagrangian approach.  
On the other hand, the next-to leading (logarithmic) contributions 
include the additional parameters other than oblique ones.  
The next-to leading contributions do not vanish in general except 
for a few cases, so that they may be useful for the indirect exploration 
of New physics by combining with data from future $e^+e^-$ colliders. 
One of the exceptional cases is that with $\cs^2 \ab \sim 0$ and 
$\mh \sim \mG \sim \mA$, which corresponds to MSSM in large $m_A$ limit. 
In these cases, both the leading and the next-to leading contributions 
simultaneously vanish and the model then becomes a decoupling theory 
as expected.
Otherwise, the non-decoupling effects on $\delta^{(1)}$ exist and 
can become large with keeping the constraint 
$\delta^{(0)} \propto \alpha_f T \sim 0$.  
One example for such cases may be $m_{H^\pm} \sim m_{H^0}$ and 
$\alpha - \beta \sim \pi/2$. 
Then the values of $m_{A^0}$ and $m_{h^0}$ can be taken 
freely with keeping $\al_f T \sim 0$.    
Actually we have numerically found that in these cases 
there can be the relatively large deviation from SM, 
which amounts to 2-3 \% (see Fig.2) at $\sqrt{s} = 1$ TeV 
for large $m_{A^0}$ but within the constraint from the perturbative unitarity. 
Therefore the non-decoupling effects by heavy Higgs bosons  
on this process can become large to some extent, 
so that they may be constrained 
by the data from future $e^+e^-$ linear colliders.

\vspace{1cm}
\noindent
{\large \em Acknowledgments}

The authors would like to thank Y. Okada for valuable discussions, 
K. Hagiwara for useful comments.


\end{document}